# Digitalization and virtual assistive systems in tourist mobility: evolution, an experience (with observed mistakes), appropriate orientations and recommendations

Bertrand David[0000-0003-0730-5125] and René Chalon[0000-0002-1129-3908]

Université de Lyon, CNRS, Ecole Centrale de Lyon,
LIRIS, UMR5205, F-69134, Lyon, France
`Bertrand.David@ec-lyon.fr, Rene.Chalon@ec-lyon.fr`

**Abstract.** Digitalization and virtualization are extremely active and important approaches in a large scope of activities (marketing, selling, enterprise management, logistics). Tourism management is also highly concerned by this evolution. In this paper we try to present today's situation based on a 7-week trip showing appropriate and shame situations. After this case study, we give a list of appropriate practices and orientations and confirm the fundamental role of User Experience in validating the proposed assistive system and the User Interfaces needed for client/user satisfaction. We also outline the expected role of Metaverse in the future of the evolution of this domain.

## 1    Introduction

Tourist trips have always needed supportive assistance [1], which was initially up to 10-15 years ago only physical and paper-based and mainly prior to the trip. The evolution of assistive systems is permanent and will continue in the future. We can mention the evolution of physical agencies to virtual technologies through the emergence of Internet [2]. The evolution of one-shot contact to continuous access to appropriate information, contextualization by geolocalization, and the internet of things must be emphasized. In this new context, User Interfaces and human behaviors have become the main factors of success of this evolution, characterized by utility, usability, and user satisfaction. Due to continuous complexification, these factors are not easy to guarantee.

Digitalization and virtualization are extremely active and important approaches in a large scope of activities (marketing, selling, enterprise management, logistics) [3]. The objective is to replace physical documents and face-to-face contact by digitalized documents shared worldwide and by remote virtual access to their exchange and manipulation. Human face-to-face is replaced by independent manipulation of these documents (data supports). Tourism is also significantly concerned by this evolution, not only in the trip preparation stages, but also and mainly in mobile situations during trips.

For this paper we had two possibilities of presentation, either a general (abstract approach) followed by case studies or an opposite one, a case study showing the most frequently observed misbehaviors and bad situations followed by local correction suggestions and only, at the end, a trial to synthesize and give an open list of recommendations. We decided to apply this second approach, after a brief presentation of the state-of-the-art of digitalization in general and its rapid synthesis in the tourism industry.

Our objective is to present first a case study based on a 7-week trip in the preparation, execution, and consolidation stages. The objective is to show appropriate as well as inappropriate behaviors and management situations. Based on this experience and a general appreciation, we draw up a digitalization schema for the tourism industry and a list of recommendations and suggestions to consider.

## 2  State-of-the-art

To address this issue, it seems interesting to describe the state-of-the art-in three fields that structure our approach: An informal view of digitalization of tourist activities, digitalization approaches and principles in various domains of use, and User Experience principles as the approach to observe and evaluate user behaviors in various situations.

### 2.1  Digitalization of tourist activities

We are not able to define once and for all the entire scope of tourist activity digitalization. As such, we propose to express it in a more informal manner.

**The influence of digitalization on tourism.** Digital technology has a profound influence on tourist economy everywhere on the planet. Trends are digital for cities, tourist offices, and travel professionals. To offer an ever more unique and impactful experience, there is no shortage of technological means. Cities are investing in tactile and interactive media [4]. Cities and communities are modernizing and investing in interactive technologies to respond to the digital transformation of uses. In tourist offices it is no longer rare to take advantage of dynamic displays and other tactile supports that enrich and facilitate the presentation of expected vacancy locations to potential clients [5]. The clever combination of broadcast screens, touch terminals, and interactive media makes it possible to create an augmented reception area: a modernized place at the cutting edge of digital technology with real added value for tourists seeking information. The positive consequences are numerous: in addition to improved communication within tourist offices equipped with digital displays, visitors benefit from a better quality of service and information than before. Multiple language facilities are also available. Large amounts of data can also be collected, analyzed, and then used with the aim of improving visitor satisfaction: performance of advisers, attendance, visitors' impressions, and their recurring requests, etc. However, in cities it is not just tourist offices that are equipped with the latest interactive technologies. Museums, shopping malls, fairs, festivals, and other tourist sites are also equipped with them in order to offer travelers an unforgettable and rich experience [6, 7].

**The tourism industry is reinventing itself.** Technological developments have caused new forms of tourism to emerge:

- **m-tourism**, also called mobile tourism, is the action of booking holidays directly via a mobile or touch pad [8];
- **Social tourism** consists in choosing your stay based on social networks [9];
- **E-tourism** is the association of web and tourism giving rise to all kinds of digital experiences [10].

To meet the growing expectations of increasingly demanding tourists, comparative sites and tools have multiplied and make it possible to create ever more personalized trips.

**The travel agency model is changing.** Another proof that tourism is directly impacted by the digital transition is that traditional travel agencies, where customers used to go to book their stays, are threatened. The Internet has seen countless tourism stakeholders flourish in just over a decade, whether they are online agencies, flight comparators, hotels, or destinations. However, all is not lost for physical agencies. Various studies have shown that 8 out of 10 tourists go on the internet to search for their holidays, but only 1 out of 3 actually book from a screen [11]. Many tourists prefer to speak with an adviser, for example to be reassured. This is where traditional agencies can make their mark by installing interactive screens, touch screens, and digital signage technologies on their premises to offer visitors a more realistic view than ever before. The immersive and interactive travel agency of the 21st century is already here and continues to grow via the Metaverse approach [12].

### 2.2 Digitalization principles

Digitalization is the transformation of current working principles in the physical world to a new world that is more or less digital and virtual [13]. This evolution started at least 15 years ago and has progressed constantly rather than linearly. It can be applied to marketing, to CRM (Customer Relationship Management), to production, to public services, to logistics and so on [14]. While information-based services can be totally digitized, physical object-based activities, production, and manipulation mix physical and virtual activities with the Internet of Things (IoT) approach [15]. Information digitalization can ensure substantial improvements to data management, thanks to the quality obtained, process time reduction, and associated costs.

Commonly, digitalization is presented as a 3-step process starting with Digitization; transformation of physical documents (texts, drawings, etc.) to digital ones, which can be more easily transported and modified. The next step is known as Digitalization, which is characterized by the presence of manipulation tools (for these digital documents) allowing these data to be shared and presented on different websites. Their trans-portation between different locations of use, mainly by computer networks and commonly accessible stores, either in specific servers or increasingly more in no-located cloud infrastructures, are thereby facilitated. The last and main step is known as Digital Transformation, which is concerned by the proposals of new organizations considering previously explained elements, leading to what is generally called Informatization [16].

In this way, economic, social, and cultural relationships and barriers are reconsidered and profoundly modified. As stated earlier, methodological processes are not the same as the scope of actions is different. Nevertheless, generic digitalization processes exist and can be used either directly or after adaptation to the field concerned.

Our paper does not focus on the design process of digitalization but rather adopts the opposite approach. We observe in it an existing digitalized environment (system) from the user/customer point of view and then suggest appropriate evolutions and improvements. This is why we give only a brief survey of possible approaches.

Globally, we can define digitalization as a transformation process, the goal of which is to propose: new practices, a new ecosystem, new stakeholders, a new business model, new management in order to provide better customer services.

For this purpose, more or less elaborated processes are proposed. We shall mention only two: the McKinsey 7S model (**Fig. 1**) [17] giving the main components of an appropriate digitalization approach with a Strategy, a Structure, Supportive Systems and managing Styles, working Staff, and members' main Skills manipulating Shared Values.

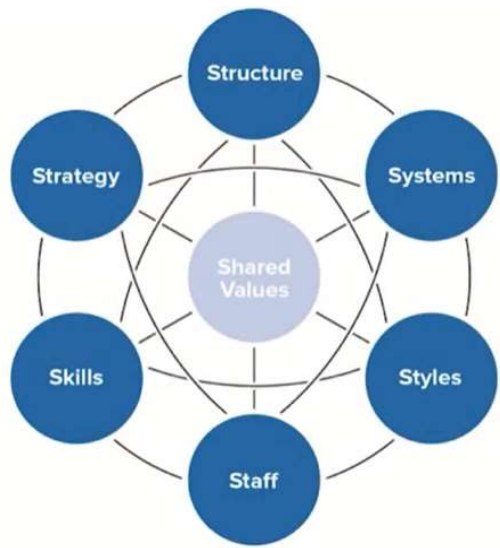

**Fig. 1.** McKinsey 7S model [17]

More fine modeling is proposed on The Big Picture of Digital Transformation (**Fig. 2**) [18], synthesizing all aspects, elements, and methods. These approaches organize design and development stages, which are not our objective in our approach, the goal of which is to recall an experience and formulate appropriate evolutions.

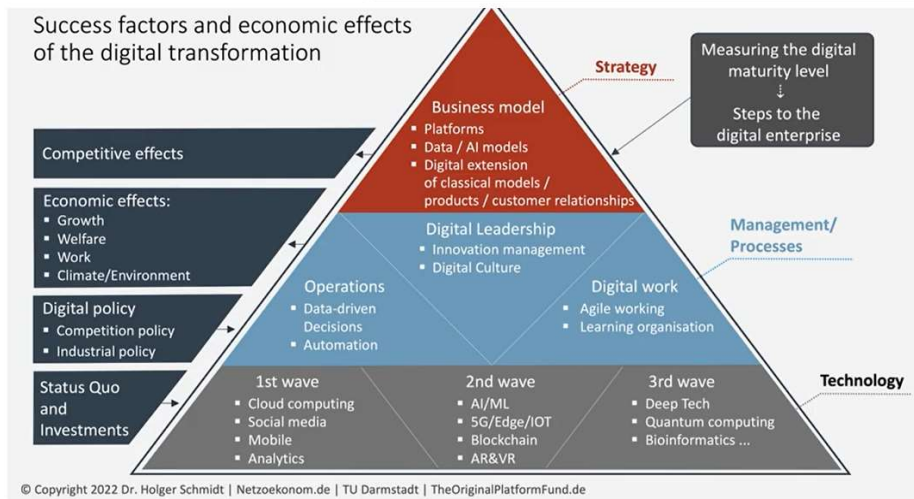

**Fig. 2.** Digital Transformation - The Big Picture (Edition 2022) [18].

### 2.3 User Experience (UX)

The term UX (User eXperience) refers to the quality of user experience in any interaction situation. UX describes the overall experience felt by the user when using an interface, a digital device or, more broadly, when interacting with any device or service. UX is therefore to be differentiated from ergonomics and usability.

It was Donald Norman, in the 90s [19], who first used this term "user experience" that would then go on to enjoy the success we know. Donald Norman recalls in a very short and instructive video [20], the origin of the term "UX" and the vision he has of it today, an open vision that makes us challenge certain ideas. Donald Norman and Jacob Nielsen in their NNGroup [21] continue to promote and evangelize this concept with a series of conferences and videos. User Experience can be seen as design and test methods conducted by professional designers, but also as final user tests in working conditions of the application or system [22].

Concerning the digitalization process, we choose to use User Experience as final user of our case study situations and collect positive and negative observations. However, our real position as computer and UI specialists allows us to formulate later some observations and suggestions for evolution of different elements (services, applications) that we used during our case study.

## 3 Long-term case study

Our case study is related to a relatively long tourist experience (7 weeks) organized initially from France and taking us to the United States for a trip covering California, Nevada, Arizona, New Mexico, Utah, and Colorado. The objective was to visit several National Parks and museums and to participate in local activities organized during our stay in different locations. We relate the main tourist services used and describe their level of digitalization and our experience (positive or negative). Main characteristics of our trip were: 2 successive rental cars used, 3600 miles covered, 43 nights in 18 hotels, 5 State Parks visited, participation in 2 major events, 5 museums visited, etc.

The trip itself will not be described, as it is not in the scope of our study. However, its geographical print is presented in **Fig. 3**.

For reasons of ethics, we decided not to publish the names of the companies and enterprises we used and evaluated. There are two reasons for this: first, a company, on which our observation is based, recognizes the described behavior without stigmatization and appropriately modifies the identified services. That is an interesting result. Second, if a company, which was not observed by us, is concerned it could be in the same case and updates its services, the impact will be greater.

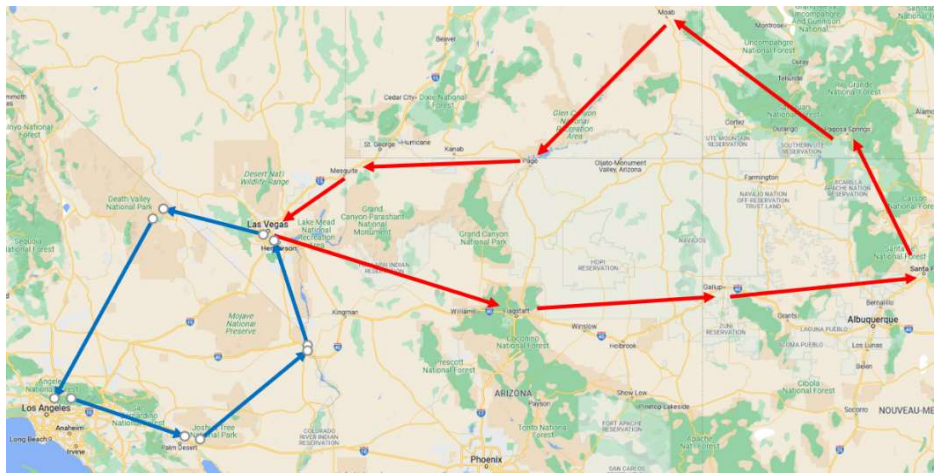

**Fig. 3.** Two-loop trip (first in red, second in blue).

### 3.1   Trip preparation

The first major activity carried out well before the trip is its preparation for which the physical and digital worlds are appropriately explored. In the physical world, discussions with friends, colleagues, and tourism professionals, consultation of books, flyers, etc. are the practices most commonly deployed. In the digital world, multiple websites are available with search engines, while social networks, such as WhatsApp, Messenger, Twitter, Facebook, and so on, make it easy to find the information you need. In this step, exchanges are open, without commitment, and tools work appropriately. Visa-Platinum and other high-level credit cards also provide a source of inspiration, making it easy to organize quality trip programs tailored to your needs.

## 3.2 Flight finalization

The next step is transportation finalization with precise timing decisions and financial agreements. In-city travel agencies have ceased to be the only solution today, as the vast majority of travelers use Internet. A variety of formulas are possible: fly only or fly & hotel packages, and so on.
In this context, appropriate digitalization is mandatory, with integrated support during and after the trip. It is important to share the appropriate data model, managing all necessary data, as well as a User Interface between provider and client as far as possible in real-time.

We shall now describe various problems that may occur. When booking a flight, the website may crash for a variety of reasons, such as network disconnection, evolution of availability of chosen option or flight fare. The wrong solution is to ask the user to restart the process. Usually, websites are able to store the data collected for this partially processed booking and continue via an alternative solution as soon as possible (using a mobile application, phone call with help by an assistant, etc.) based on a multichannel approach, in order to finalize this sales action securely and as quickly as possible.
We had the opportunity to benefit from the possibility of finishing our purchase after an initial website crash with a flight booking operator.

After this initial action, it is important to remain in contact with the operator for the following official procedures (registration, complementary baggage payment, options, etc.), as well as for more personal ones (trip modification, cancelation, etc.). These contacts must be as reactive as possible. Unfortunately, it is not always easy to obtain an answer quickly. As an example, we can mention an answer received a month and a half after our trip concerning additional baggage costs. Some messages related to transportation reservation codes are managed appropriately and in accordance with time constraints, while others are unfortunately treated more randomly.

We observed several difficulties during check-in, for example visa verification, COVID certificates, and vaccinations were impossible to be done on-line, making it necessary for them to be checked at the airport.

For post-trip actions an appropriate behavior is expected: reactivity and a rapid answer, also for complaints, with a two-step approach: confirmation of complaint reception, then later, but not too late, the final answer. Reactivity to emails is a huge problem as answers are returned far too late for the problem in question. Complaints are all too often processed after a long wait, if at all.

Aviation transport introduced at an early stage digitalized services to replace the tourist agency approach. The search for trips and flights are appropriately supported, as well as payment.

Unfortunately, this is not the case for more specific issues, where it is possible to contact companies only by phone (with cost and communication problems). The majority of emails sent by companies are no-respondable, i.e. it is impossible to use them to ask a question, point out a problem or send a complaint. One solution is to use pre-established formulas, but these are often limited to a closed list of situations (FAQ), unfortunately, without open-ended ones.

### 3.3 Car rental experience

Car rental is an important aspect, particularly for trips to other countries or continents. Tourist agency physical reservation is progressively being replaced by digital reservation, making it necessary to contact directly via internet specific car rental servers or a generic server working for multiple car rental companies or to use related services proposed by airlines, hotel reservation servers or high-level credit card bank services.

In most cases the process is easy: choice of location, rental duration, car category, and complementary services. On arrival in the car rental agency, the document (voucher) is presented, together with an ID, a credit card, and a driving license. The employee draws up the rental contract, charges the credit card, and asks the client to sign it. The client can then go to the rental carpark to choose the car in the chosen rented category, check it with the employee, and indicate any problems observed (often car-body impact).

Unfortunately, we did not go through this generic process for our first car rental. Our reservation process started with the bank and its high-level credit card service. This service offered us a rental company and a rental price. On arrival at the rental agency, the employee verified our ID, driving license, and charged our credit card, without any contract to sign, before giving us the car key and indicating the carpark. We left the carpark without any car state checking or document, but with an important statement: we are "PAPERLESS", you will receive an email with all appropriate information. Several days later, we received the promised email in which we could read all the car-body impacts declared by previous rental users. However, we also observed that the rental period was not ours (as indicated in the voucher). It was modified (3 days shorter) at the same price. We tried to find out what had happened but the car rental company's email was non-answerable. We then contacted our bank that had proposed us this rental and their answer was "go back to the rental agency to modify this problem". However, we were 200 miles from the initial location. The second email stated: "the rental company is not able to establish more than a one-month rental (our rental was 32 days), they will send you on the last day of your current rental period a new contract covering the missing period of 3 days". We went along with this, but what actually happened was totally different. We never received the additional email with the new contract and on the last day of the first period, our bank account was charged for the entire amount. Three days later, when we returned the car we went through the same process. The employee asked for our car key, checked the car, and told us everything is OK, you will receive an email, as we are "PAPERLESS". We never received this email, but our card was charged for an extra $900. We tried to inform our bank and both the US and the French car rental agencies. For two weeks we received no answer. Two weeks later our bank account was credited for two sums adjusting the cost of the first and second periods, and we received a partial explanation from the car rental agency stating the impossibility to rent a car for more than one month.

To wind up this case, two weeks later we received from a higher-level manager of the US car rental agency the following and really surprising email (**Fig. 4**)

> *Please note that all our vehicles are carefully cleaned after every rental. This is already factored into our rental prices. After verifying to one of our associates at our Las Vegas Int Airport location, it has been advised that the vehicle you returned required a special cleaning due to it being dirty. Therefore, cleaning the vehicle took considerable time and expense and for this reason you have incurred the additional charge for cleaning. based on your last contract 9024360913, vehicle was returned on AUGUST 30 2022. The charge on our end is $916.09 on AUGUST 31 2022.*
>
> *But it was removed and the only charge appearing in this contract is for $245.96 which is for your remaining rental days, from AUGUST 26-30 2022. I hope this information clarifies your inquiry.*

**Fig. 4.** Last e-mail from US car rental agency.

Of course, this observation about "special cleaning due to it being dirty" was unfounded as we were two in the car (without kids), never ate inside and never drove off official roads. The check-out employee did not indicate any problem.

Fortunately, our second car rental experience went smoothly as per the normal procedure before, during and after rental, allowing us to verify that our first experience was a counterexample, as well as helping us restore our confidence in this kind of service. In the last section of this paper, we will summarize our suggestions for appropriate behaviors of different digitalization stakeholders.

We finish with an open question: How would you react if you saw the following message on your dashboard, see **Fig. 5**.

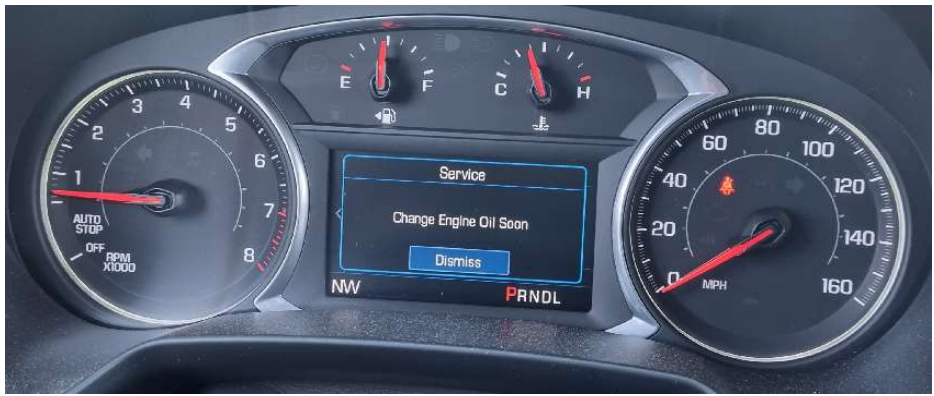

**Fig. 5.** Change Engine Oil Soon.

### 3.4 Hotel reservation management

Hotel reservation is another important aspect in tourist trips. Assistive Systems for hotel reservation are progressing accordingly. After physical reservation via a tourist agency, phone calls, and direct snail mail contact with the required locations, websites now increasingly propose appropriate services via specialized, individual websites, via more or less aggregated solutions by hotel chains (homogeneous Best Western, Hilton, etc.), heterogeneous solutions (Choice Privileges) or via totally generic solutions such as Hotels.com or Booking.com. The level of services seems appropriate, with in-depth selection tools, appropriate reservation services, and the possibility to cancel or modify reservations up to 24 hours beforehand. Multichannel communication is also appropriately supported.

We encountered a problem related to the COVID and post-COVID periods: no information as to the type of breakfast (English, French or grab-n-go) or how the rooms are or are not cleaned (only on request to avoid housekeeper entry, only towels and hygienic products provided, classical cleaning). This information was either given during check-in or observed at the end of the day when the room was not cleaned, and we wanted to know why. Another problem was with refunding delays when we decided, for an objective (pool out of order or grab-n-go breakfast) or for a personal reason, to shorten our stay. As the hotel debited the totality of our stay on our arrival, the bank took 10 to 15 days to refund us.

### 3.5    Bank role

The role of banks in mobility is mainly Credit Card management-oriented. These cards offer an implicitly limited amount for payment and withdrawal, which can be more or less easily modified. Self-modification on the customer space is an appropriate solution, but not necessarily totally generalized (as the maximum amount fixed by the bank cannot be exceeded).

Greater modification possibilities necessitate participation by bank employees, which is not always easy and is mostly dependent on their availability during vacation periods and bank opening hours (not necessarily compatible with worldwide tourist locations). The main channel used is still the phone, which is not necessarily appropriate for overseas calls. Multichannel and appropriate reactivity are the objectives.

### 3.6    Insurance company participation

In the tourist industry, mobility insurance companies offer at least assistance and rescue services, but also services for car rental, hotel and transportation reservation. In these cases, reactivity and continuity of the customer relationship are essential. The problems of rental car breakdown or accidents need appropriate reactivity and a relationship with the rental car company, which is not necessarily always the case. For various reservations (hotel, trip, etc.), it is not easy to offer the same level of services as that offered by specialized providers.

### 3.7    In-site access tools and assistance

During our trip we observed several specific digitalized services for particular situations. We noted a timed entry reservation that was required to enter the park even if you already have a pass. Annual passes cover the payment of entrance fees, so annual or senior pass holders only need to pay the additional $2 reservation fee. This timed entry

reservation is used in several US National Parks in order to manage rush hours of access to the parks during the summer season (**Fig. 6**). A website is devoted to the distribution of timed entry tickets in relation to the number of potential visitors [23].

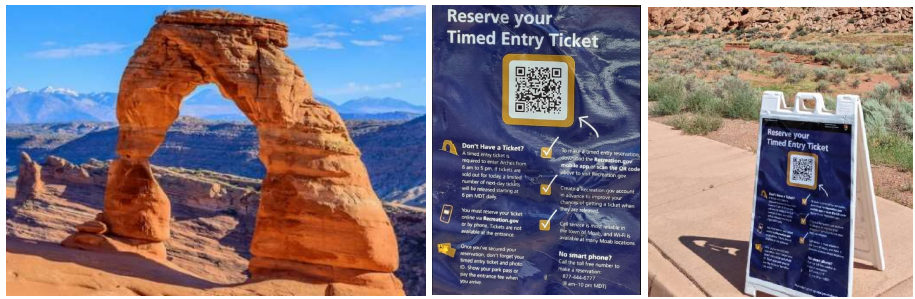

**Fig. 6.** Arches National Park access control.

We found similar services for museum access management for the Getty Villa and Getty museum in Los Angeles. Comparable services were also used to manage entry and car parking at the Gallup Inter-Tribal Indian Ceremonial and Santa Fe Opera. Generic tools such as Uber-Eat, etc. can easily be added to the above.

### 3.8 Chatbot solicitation

Textual-and speech-oriented chatbots are helpful if they work correctly. Unfortunately, this is not generally the case. In particular, in situations in which the chatbot tries to answer open-ended situations, its capacity of understanding is limited. Often it is unable to contextualize correctly the question (the problem).

An approach in which the chatbot leads the interaction is more success-oriented. In other words, the chatbot manages the interaction by proposing a list of situations (FAQ) and asking users to choose one. The hierarchical structure can then be used to manage the exchange process to discover the contextualized situation and find the appropriate answer. Of course, this approach can prove too complicated if the selection tree is too large or deep.

An interesting solution would be to combine the chatbot with human interaction, where the chatbot orients the problem formulation and the human answers the targeted problem, if the chatbot is unable to understand and answer.

### 3.9 Human implication

Human implication in the digitalized process is a major contribution that must be studied and answered appropriately. Of course, for cost reasons all measures are taken at present to do away with human participation as it is both time and cost consuming. However, the latter participation is also the ultimate solution for a large number of user situations that are not correctly treated by chatbots. A speedy response to their questions is essential to ensure that users do not give up.

### 3.10 Generalization to other fields

The previous list of observations was based on our trip experience. Of course, we are able to add other experiences, not directly related to tourist mobility but nevertheless characterizing problems poorly treated during digitalization.

One of the official French websites aims at managing delivery of all official documents such as driving licenses, ID, passports, and vehicle identification. This service is totally digitalized, with no physical counter. There is just a phone number on which a robotized human gives only stereotyped answers as a robot. On the website you can explain your problem for predefined situations only (problem FAQ) and no open-ended questions can be formulated. After submitting your question, the system acknowledges it and informs you that you will receive an answer within 48 hours. Unfortunately, reality is a little different, as you will normally receive a first, very short, answer after one week.

### 3.11 Non-observed services and situations

Unfortunately, during our trip we did not observe any contextual geolocalized services based on the IoT (Internet of Things). These services combine virtual services with in-the-field located objects such as opening closed roads based on special authorization transmitted from the assistive system to remote objects via a personal device such as a smartphone.

Another example is dynamic lane management making it possible to open or close a lane to a category of vehicles. The lane can be closed to private vehicles and dynamically devoted to buses or emergency vehicles [24]. The objective is to devote these lanes dynamically only when buses are present and to leave all lanes open to general traffic when buses are not present. This allows management of general traffic speed. The main technologies used are: a Location-Based Service integrating bus detection sensors; an intermediation platform collecting sensor information and determining dynamic bus section activation and deactivation; in-the-field infrastructure and/or embedded vehicle interface receiving instantaneous information on selected situations (**Fig. 7**). From the HCI point of view, it is important to indicate the present situation on in-the-field indicators, as well as on the screen in the vehicle.

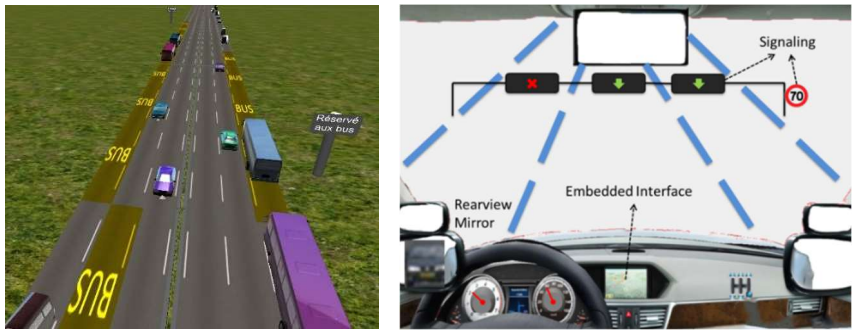

**Fig. 7.** Dynamic lane management [24].

We did not meet Metaverse services, which are still in the experimentation stage, and the objective of which is to mix reality and virtuality based on VR (Virtual Reality), AR (Augmented Reality), and Mixed Reality [25], using appropriate devices. Smartphone-based applications are the lowest configuration available.

## 4  Assistive system structures

We can summarize all assistive services described (and used) during our trip as follows: Airlines, Rental companies, Hotels, Banks, State parks, Museums, Event structures (Santa Fe Opera, Gallup Intertribal Ceremonial, restaurants, etc.), GPS, food delivery, etc.

These are either elementary services or more sophisticated multi-services. They are made up of different aspects, namely: user interface, working data, one or more services (algorithms working on data), and remote access mechanisms using internet or other network services.

From an integration point of view, we find either elementary configurations, limited to one issue, or more or less integrated multi-issue configurations. It is clear that this is not an integrated system but only a partially integrated approach. Naturally, banks are connected / interfaced with a large majority of proposed services, mainly for payment needs. Banks also propose related activities that require integration or at least data exchange with other applications (hotel reservation, rental cars, etc.). Just like banks, airlines also propose related services such as hotel reservation, rental cars, etc. Hotel reservations are more or less homogeneous, either individual or hotel chains: homogeneous (Hilton, Best Western, etc.), heterogeneous (Choice Privilege, etc.) or totally generic (Booking, Hotels, etc.) with appropriate management of reservations, modifications, and so on. Individual services are also used for State Park access, museums, and even structures.

It appears that the notion of silo continues to be an integration model, with well-defined main operations able to exchange easily and rapidly. As we will see later, some less frequently used operations require more or less major improvements.

Inter-silo relations are in the same situation with appropriate major exchanges and the necessity to improve less frequently used operations.

Total integration seems impossible to obtain, as the global system is not closed but rather open-ended, able to receive (each day) new services and applications supporting new visions of digitalized / virtual tourist activities.

Introduction of a new digitalized service can be either independent or need to collaborate with other services already present. These services must provide an API (Application Programming Interface), a standardized approach for their inter-connection.

As we can see in **Fig. 8**, GPS is an independent digitalized service, while Bank is a more common digitalized service, mainly for payment. Collaboration between services means that a service can be used while another major service is being used, such as hotel reservation from an airline digitalized service. It would appear that integration based on systematic data standardization is not appropriate, as data used by different services are at least partially different, and their integration could be too complicated, necessitating complex exchanges between services. A mixed approach, with a local data store for specific data and a shared store in a chosen location for shared data would seem appropriate (payment data stored in a Bank service).

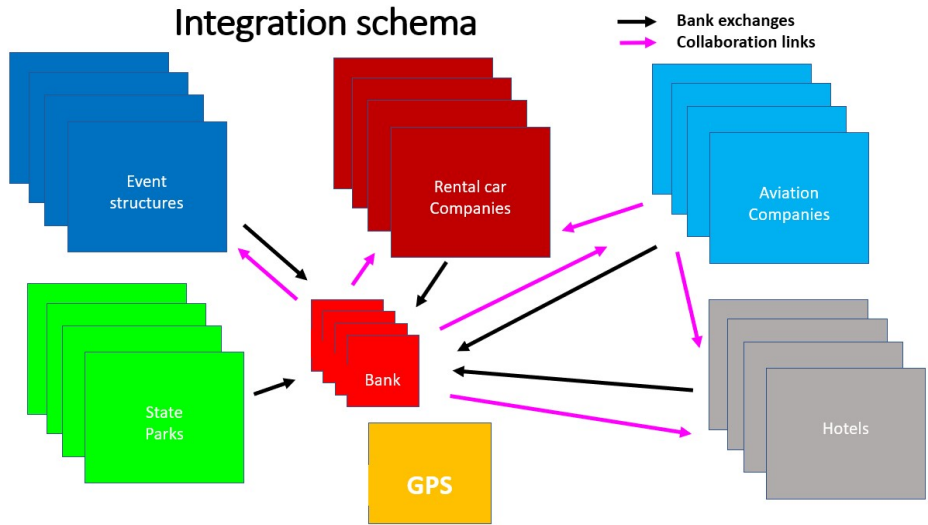

**Fig. 8.** Silo and inter-silo exchanges and collaborations.

Two technics can constitute a major breakthrough by the use of available data stored in the system. Using the big data approach [26], it is possible to extract from these data more general behaviors increasing the possibility of new services. In another direction, reuse of available data in Machine Data learning and Deep Learning approaches can add important Artificial Intelligence-based behaviors to the system [27].

## 5      Identified improvements to be considered

After our trip and the corresponding documentation of appropriate and inappropriate behaviors of digitalized and virtual services, we can now propose a list of improvements, which we have decided to divide into 3 sections:

- Application-oriented:

1. **Avoid inappropriate use of certain concepts**, such as "paperless" in those contexts in which it is mandatory to collaborate in real time and exchange persistent documents (contract validation by all partners, state of vehicle, etc.).
2. Be able to update the websites to **integrate UpToDate information** as soon as possible (COVID restrictions, change of type of breakfast, pool out of order).
3. **Avoid incompatibility between the user/consumer data model** and the in-house enterprise model (Data model compatibility – alignment), i.e. one month-limited rental which is only an in-house choice, not visible or understandable by the client.

4. Banks, hotels, and airlines, which propose appropriate behaviors in mainstream activities, need to improve **related and peripheral activities** for which additional efforts are needed, mainly with respect to the user/customer relationship. There is a need for more efforts to provide customers with explanations and support.
5. **Maintain coherence** of information during exchanges: price changes during the check-in process (small at the beginning and larger on actual payment).

- UI-orientated:

6. **Allow multi-device and multi-channel interactions** to provide continuity of service in sequential and/or parallel manner.
7. **Provide systematically answerable e-mails** easily accessible and providing answers within an appropriate time.
8. **Avoid inappropriately low reactivity**: Example - a month and more for e-mails, even for questions needing immediate answers (could I rent a car with miles?).
9. **Avoid only pre-established / selected questions** available on digitalized forms on the Internet. Open-ended questions are also mandatory.
10. **The human relationship disappears totally** with virtualization (no window / information office available). Whenever necessary, **provide a User Interface** to maintain the relationship with the user / consumer.
11. **Provide appropriate chatbots** (textual and/or vocal) with an appropriate structure generally competent or thematic with progressive question selection. If necessary, provide real-time human intervention for more detailed discussions.
12. The **chatbot** can be used as the first step, but must be replaced by the **human** each time the former fails or is not able to provide appropriate help.
13. The **human must be ready to participate** whenever the user/consumer needs him/her.

- Evolution-oriented:

14. Integration of new services / applications requires a global study with appropriate collaboration between services to determine appropriate inter-relationships of shared data based on digitalization methodological processes such as Design thinking [28].
15. Appropriate UX for all users / customers is mandatory.
16. IoT for in-city (in the field) geographically situated services
17. Metaverse as a promising evolution that is not yet sufficiently mastered and proposed.

## 6   Conclusion

In this paper we studied digitalization and virtualization of operations related to tourist mobility. We based our approach on a concrete case study of a 7-week trip to the United States of America. After an initial contextualization of digitalization and virtualization in general and, more specifically, to tourist activities, we drew up a list of experiences related to our trip covering preparation, flight management, rental car experience, hotel reservation, and other specific services such as State Park access reservation. Then, we discussed more generally the level of integration of proposed services and our user experience, which led us to elaborate a list of suggestions for appropriate behaviors that tourism-oriented digitalization and virtualization could provide in order to obtain *a priori* higher level of user satisfaction.

Our conclusions are that, globally, as the proposed digitalization and virtualization for tourist mobility are appropriate for mainstream operations such as flight management, hotel and rental car reservation and management. The majority of problems occur for more specific demands and, in particular, situations such as modifications/adjustments or requests for specific services. In these situations, the lack of possibility to communicate using multiple channels is a problem, as only a mobile phone number is systematically present, but not an email. Reactivity and speed of reply are insufficient. Chatbots are not yet fully perfected, and synchronous User Interfaces are generally missing. It is our opinion that these aspects must be improved in order to provide appropriate User Experience in these peripheral, but very important, stress-generating situations.

Total integration is not a solution. Conversely, an open-ended approach could be preferred in order to integrate new services, which can be utilized on a daily basis and would communicate easily with existing services. The modularized approach is an appropriate solution.

During our trip we did not find any contextual applications connecting the user's application with in-the-field elements (opening of a door, information readable on a precise location, management of highways for specific cars, etc.) using the IoT (Internet of Things) to deploy this approach. In the future, this technology will be increasingly deployed. Another major field of tools now being extensively investigated by many companies is Metaverse [12], an approach to virtual and augmented reality allowing an interesting, rich contextualization and in sofa sitting tourism.

Generalized use of Big data and AI technics (Machine Learning and Deep Learning) [26, 27], as mentioned, can increase the capacity of these assistive systems.

Naturally, in order to confirm these recommendations, this experiment would need to be carried out with more users and at different times of the year. These tests could reveal other problems and therefore lead to new recommendations.

It would be interesting to conduct the same kind of trip with disabled users and collect their observations, difficulties, and recommendations. This could start with a comparison between a physical and a virtual travel agency for able-bodied people and for people with disabilities, before going on to address other tourist mobility services.